\documentclass[twocolumn,showpacs,preprintnumbers,amsmath,amssymb,aps,floatfix]{revtex4}

\usepackage{graphicx}
\usepackage{dcolumn}
\usepackage{bm}

\def\vek#1{ {\rm \bf  #1} }
\def \vekk#1{ {\bm #1} }
\begin{document}

\title{Exact Keldysh theory of strong-field ionization: residue method vs 
       saddle-point approximation}  

\author{Yulian V. Vanne}
\author{Alejandro Saenz}%
\affiliation{%
AG Moderne Optik, Institut f\"ur Physik, Humboldt-Universit\"at
         zu Berlin, 
         Hausvogteiplatz 5-7, D\,--\,10\,117 Berlin, Germany}%

\date{\today}%

\begin{abstract}
In recent articles [Mishima {\it et al.}, Phys.\,Rev.\ A, \textbf{66}, 033401 (2002);
Chao, Phys.\,Rev.\ A, \textbf{72}, 053414 (2005)] it was proposed to use the
residue theorem for the exact calculation of the transition amplitude describing 
strong-field ionization of atomic systems within Keldysh theory. This should
avoid the necessity to apply the method of steepest descent (saddle-point 
approximation). Comparing the results of both approaches for atomic hydrogen 
a difference by a factor of 2 was found for the 1s, and an even more 
drastic deviation for the 2s state. Thus it was concluded that the use of the 
saddle-point approximation is problematic. In this work the deviations are 
explained and it is shown that the previous conclusion is based on an
unjustified neglect of an important contribution occurring in the application 
of the residue theorem. Furthermore, the applicability of the method of
steepest descent for the ionization of Rydberg states is discussed and an 
improvement of the standard result is suggested that successfully 
removes the otherwise drastic failure for large principal quantum 
numbers.
\end{abstract}

\pacs{32.80.Rm, 33.80.Rv}

\maketitle

\section{\label{sec:Intro}Introduction} 

The ionization process in atomic systems exposed to a strong laser field 
has attracted great interest during the past decades and its proper 
modeling remains a challenge to theory~\cite{sfa:beck05}. %
Among the numerous approximations developed to treat this problem Keldysh 
theory~\cite{sfa:keld65} possesses a prominent role. This theory was proposed 
by Keldysh 40 years ago and describes the ionization process as a transition 
between an initial electronic bound and a Volkov continuum state (adopting the
length gauge formulation). Besides 
the main approximation of the Keldysh theory, the neglect of the interaction 
of the escaping electron with the long-range Coulomb potential (in the case 
of a neutral atom), two additional simplifications were made in~\cite{sfa:keld65}: 
(i) the method of steepest descent (MSD)~\cite{gen:hass99} (saddle-point 
approximation) for performing an occurring contour integral, and (ii) the 
assumption of a small kinetic momentum of the escaping electron. 
The Keldysh approximation and variants of it are also very popular, because 
they are the basis for methods that predict strong-field 
ionization rates also for heavier atoms and molecules (see, e.\,g., 
\cite{sfa:reis92,sfm:kjel05b}). 

Recently, there has been proposals to obtain an exact Keldysh theory by 
removing the two additional simplifications (i) and (ii). It was especially 
suggested to avoid approximation (i) by solving the occurring contour
integrals with the aid of the exact residue theorem (RT) instead of the 
MSD~\cite{sfa:mish02,sfa:mish02b,sfa:chao05}. It was shown that the RT method  
yields a two times larger transition amplitude for the 1s state of a 
hydrogen-like atom and, as a consequence, a four times larger ionization 
rate. Furthermore, the ionization rate of the first excited (2s) state
obtained in~\cite{sfa:chao05} when applying the RT differs 
significantly from the MSD result. In view of the popularity of the MSD 
approximation for treating strong-field problems like ionization~\cite{sfa:milo06} or 
high-harmonic generation~\cite{sfa:figu00} this is of course a very important result. %
This has motivated the present study in which a careful reinvestigation 
of the RT and the MSD is performed (Sec.~\ref{sec:Theory}). It is shown 
that the application of the RT as proposed 
in~\cite{sfa:mish02,sfa:mish02b,sfa:chao05} contains an unjustified neglect of
the contribution of one integral and that it is this omission which is the main source for the 
previously reported deviation between the RT and the MSD results. Therefore, the 
MSD provides in fact more reliable results than the (incomplete) RT approach.  
It is furthermore discussed that the MSD fails for Rydberg states and a correction 
to it is proposed. The conclusions of 
Sec.~\ref{sec:Theory} are supported with the aid of a numerical study in 
Sec.~\ref{sec:Test}.

\section{\label{sec:Theory}Theory} 

\subsection{Transition amplitude}

In order to provide the basis for the subsequent discussion and to introduce 
the notation a brief summary of the Keldysh theory is given that follows 
closely the one described in Appendix A of~\cite{sfa:grib97}.  
The total ionization rate of a one-electron atomic system with the electron 
binding energy $E_{\rm b}$ in the harmonic laser field 
$\vek{F}(t) = \vek{F} \cos\omega t$, with the period $T = 2\pi/ \omega$, 
can be expressed as the sum over $N$-photon processes (atomic units are used 
throughout this work)
\begin{equation}
W = 2 \pi \int \frac{d^3 p}{(2\pi)^3} | A(\vek{p}) |^2 \hspace{-1mm} \sum_{N = 
N_{\rm min}}^{\infty}\hspace{-1mm} \delta( E_{\rm b} + \frac{p^2}{2} + U_p - N \omega)
\label{W}
\end{equation}
where $U_p = F^2/(4\omega^2)$ is the electron quiver (ponderomotive) energy 
due to the field. The transition amplitude $A(\vek{p})$ can be calculated using
\begin{equation}
 A(\vek{p}) = \frac{1}{T}\int\limits_{0}^{T} d t 
              \int d^3 r\, \Psi^{*}_{\vek{p}}(\vek{r},t) V_F(t) \Psi_{0}(\vek{r},t)
\label{ApEx}
\end{equation}
where $\Psi_{0}(\vek{r},t) = e^{i E_{\rm b} t} \Phi_{0}(\vek{r})$ is the 
wave function describing the initial electronic state in the atomic potential
$U(\vek{r})$. Therefore, $\Phi_{0}(\vek{r})$ fulfills the stationary
Schr\"odinger equation  
\begin{equation}
\left[ -\frac{1}{2}\nabla^2 + U(\vek{r}) + E_{\rm b} \right] \Phi_{0}(\vek{r})
= 0 \quad .
\end{equation}
The interaction with the laser field is given in length gauge by 
$V_F(t) = \vek{r}\cdot\vek{F}(t)$. Finally, the 
Volkov wave function~\cite{sfa:volk35} $\Psi_{\vek{p}}(\vek{r},t)$ satisfies
\begin{equation}
i \frac{\partial \Psi_{\vek{p}}}{\partial t} = 
\left[ -\frac{1}{2}\nabla^2 + V_F(t) \right] \Psi_{\vek{p}},
\end{equation}
and can be explicitly written as
\begin{equation}
\Psi_{\vek{p}}(\vek{r},t) = 
\exp\left[ i \vek{r}\cdot\vekk{\pi}(t) - \frac{i}{2}\int\limits_0^t
  \vekk{\pi}^2(t') d t' \right] 
\end{equation}
where $\vekk{\pi}(t) = \vek{p} + (\vek{F}/\omega) \sin\omega t$ is the 
mechanical momentum of an electron with the canonical momentum $\vek{p}$ 
in the field $\vek{F}(t)$. Introducing the auxiliary functions
\begin{equation}
V_0(\vek{q}) =  \int d^3 r\, e^{- i \vek{q}\cdot\vek{r}} (\vek{F}\cdot\vek{r}) \Phi_{0}(\vek{r}) =
 i  \vek{F}\cdot\nabla_{\vek{q}}  \tilde{\Phi}_{0}(\vek{q}) 
\label{Vo}
\end{equation}
(where $\tilde{\Phi}_{0}(\vek{q})$ is the Fourier transform of $\Phi_{0}(\vek{r})$) and 
\begin{equation}
S(t) =  \int\limits_{0}^{t} d t' \left[ E_{\rm b} +  \frac{1}{2}\vekk{\pi}^2(t') \right]
\end{equation}
the transition amplitude $A(\vek{p})$ can be rewritten as
\begin{equation}
 A(\vek{p}) = \frac{1}{T}\int\limits_{0}^{T} d t \:  \cos(\omega t) 
                      \: V_0\left(\vekk{\pi}(t)\right) \: e^{i S(t)} \quad .
\label{Ap2}
\end{equation}

\begin{figure}
\includegraphics[width=80mm]{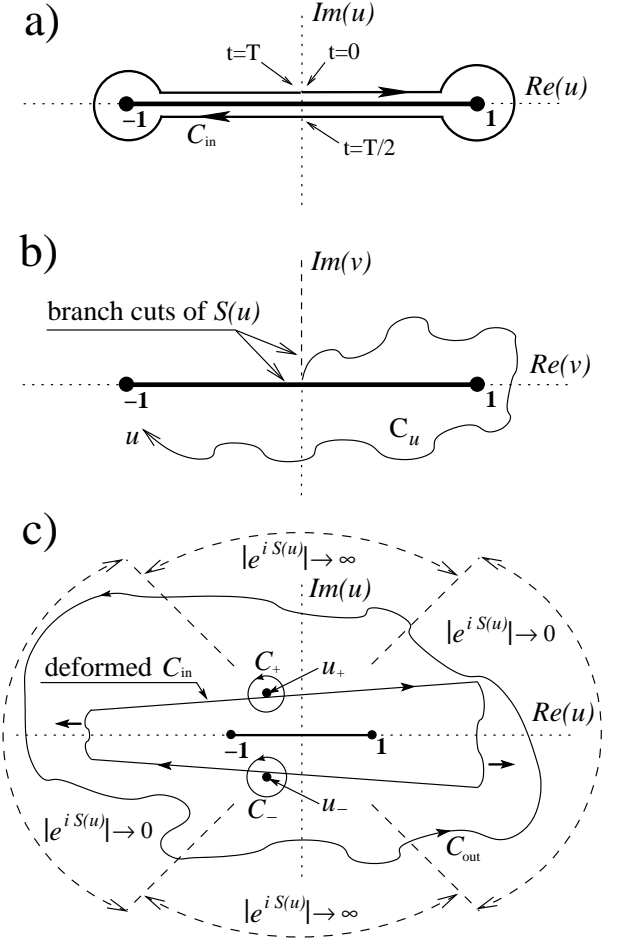}
\caption{\label{fig:ContA} 
({\bf a}) The path $C_{\rm in}$ of the contour integration for $A(\vek{p})$.
({\bf b}) The path $C_{u}$ of the contour integration for $S(u)$.
({\bf c}) The asymptotic behavior of $\exp[i S(u)]$, contours $C_{\pm}$ around 
saddle points $u_{\pm}$, contour $C_{\rm out}$, and deformed $C_{\rm in}$ used 
within MSD.
}
\end{figure}

The equivalence of $A(\vek{p})$ in (\ref{Ap2}) and $L(\vek{p})$ 
in Eq.\,(15) of the original Keldysh work~\cite{sfa:keld65} can be shown 
in the following way. The path of the integration over $t$ in (\ref{Ap2}) 
can be shifted into the complex plane by means of the transformation 
$\tilde{t} = t + i \epsilon$ where $\epsilon$ is an infinitesimally small 
positive number. Introduction of the new complex variable 
$u = \sin\omega\tilde{t} = \sin\omega t + i \epsilon\cos\omega t$ transforms 
the integration $\int\limits_{0}^{T} d t$ to one on the closed contour 
$C_{\rm in}$ which encloses the interval $(-1,1)$ (see Fig.\,\ref{fig:ContA}\,a). 
Applying the same procedure to the integral contained in the function $S(t)$
yields 
\begin{equation}
 A(\vek{p}) =  \oint\limits_{C_{\rm in}} d u  \,\tilde{V}_0\left(u\right) e^{i S(u)}
\label{Ap3}
\end{equation}
where
\begin{eqnarray}
\label{Vou}
 \tilde{V}_0(u) &=& \frac{1}{2\pi} V_0\left(\vek{p} + \frac{\vek{F}}{\omega} u  \right) \\
 S(u) &=& \int\limits_{C_u} \frac{d v}{f(v)} \left[ \frac{E_{\rm b}}{\omega} +  
\frac{1}{2\omega}\left(\vek{p} + \frac{\vek{F}}{\omega} v \right)^2 \right]. 
\label{Su}
\end{eqnarray}
The transition amplitude $A(\vek{p})$ in (\ref{Ap3}) is for $f(v)=\sqrt{1-v^2}$ 
identical to $L(\vek{p})$
in~\cite{sfa:keld65} which is also the starting point of the analysis
in~\cite{sfa:mish02}. Since the square root is usually defined to possess a
non-negative real part which can in the present context be misleading and
cause a sign error, we introduce the function 
\begin{equation}
     f(v) = {\rm Sign}[{\rm Im}(v)] \sqrt{1-v^2},
\end{equation}
which is analytical in the whole complex plane except its branch cut $[-1,1]$.
The path of integration $C_u$ specifies the clockwise path around the branch 
cut (see Fig.\,\ref{fig:ContA}\,b) starting at $v=i \epsilon$ and terminating
at $v = u$. Note, $S(u)$ is a multivalued function, so we have selected 
also the branch cut along positive imaginary axis. 

Due to the delta function in Eq.(\ref{W}) one needs to calculate $A(\vek{p})$ 
only for $|\vek{p}| = p_N = \sqrt{ 2(N\omega - E_{\rm b} - U_p)}$. Both (\ref{Ap2}) 
and (\ref{Ap3}) can equivalently be used for numerical integration to yield an exact result. 
The use of (\ref{Ap3}) provides more flexibility, since the contour $C_{\rm in}$ 
can be deformed in a convenient way. 

There exist two special points $u_{\pm}$ (\ref{upm}) in the complex plane $u$. They are 
simultaneously the saddle points of $S(u)$ and poles of $\tilde{V}_0(u)$. 
In~\cite{sfa:keld65} Keldysh has used MSD to approximate $A(\vek{p})$. 
In~\cite{sfa:mish02} the authors have proposed to use RT~\cite{gen:hass99} for
an exact calculation of $A(\vek{p})$. For the 1s state of hydrogen-like atoms 
the expression for $A(\vek{p})$ obtained in~\cite{sfa:mish02} is larger than
that of Keldysh using MSD by exactly a factor of two, provided the small
$\vek{p}$ approximation is consistently used or omitted in both the RT and the 
MSD approach. As is shown below, the disagreement is a consequence of a wrong 
assumption made in~\cite{sfa:mish02}. In fact, for excited states of
hydrogen-like atoms the therein proposed approach may lead to drastically wrong 
results. 

For spherically symmetric bound states of hydrogen-like atoms with principal 
quantum number $n$ the function $\tilde{V}_0(u)$ can be presented (see Appendix B) as
\begin{equation}
 \tilde{V}_0(u) = \frac{g_{+}(u)}{(u - u_{+})^{\nu}} =  \frac{g_{-}(u)}{(u - u_{-})^{\nu}},
\quad \nu = n + 2
\label{Vo_rep}
\end{equation}
where $g_{\pm}(u) = \tilde{V}_0(u)(u - u_{\pm})^{\nu}$ is an analytical 
(and, possibly, slowly varying) function in the vicinity of the special 
points $u_{\pm}$. Both procedures (MSD and RT) are considered in this work for 
general $\tilde{V}_0(u)$ having poles of order $\nu$ at $u = u_{\pm}$.

\subsection{The residue theorem}

Since for $|\vek{p}|=p_N$ the function $\tilde{V}_0(u)\exp[i S(u)]$ is analytical 
in the whole complex plane except the branch cut
$[-1,1]$ and the poles $u_{\pm}$, (\ref{Ap3}) can be modified using
\begin{equation}
 \oint\limits_{C_{\rm in}} =  \oint\limits_{C_{+}} + \oint\limits_{C_{-}} - \oint\limits_{C_{\rm out}}
 \label{CEq}
\end{equation}
where $C_{\pm}$ are contours around $u_{\pm}$ and $C_{\rm out}$ is a contour 
enclosing $(-1,1)$ and $u_{\pm}$ (see Fig.\,\ref{fig:ContA}\,c). The integrals $A_{\pm}$ 
along $C_{\pm}$ can be calculated using RT which yields
\begin{equation}
 A_{\pm} =  \frac{2\pi i}{(\nu-1)!} \lim_{u \rightarrow u_{\pm}} \frac{d^{\nu-1}}{d u^{\nu-1}} 
\left[ g_{\pm}(u) e^{i S(u)} \right] \quad.
\label{Apm}
\end{equation}
With the knowledge of the integral along $C_{\rm out}$,
\begin{equation}
 I_{\rm out} =  \oint_{C_{\rm out}} d u  \tilde{V}_0\left(u\right) e^{i S(u)},
\label{Iout0}
\end{equation}
the value of $A(\vek{p})$ can be calculated using (\ref{CEq}) as
\begin{equation}
 A(\vek{p}) = (A_{+} + A_{-}) - I_{\rm out}.
\label{Ap4}
\end{equation}
In~\cite{sfa:mish02} the value of $I_{\rm out}$ is implicitly assumed to be
zero. However, a simple analysis shows that there are no reasons for such an
assumption. Indeed, for $u = R e^{i\theta}$ with $R \rightarrow \infty$, one
finds [see Eq.\,(\ref{expiSu_as})]
\begin{equation}
e^{i S(u)} \rightarrow e^{- (U_p/\omega) R^2 \cos (2\theta)} e^{- (U_p/\omega) R^2 \sin (2\theta) i}.
\label{limEiS}
\end{equation}
Since $\tilde{V}_0(u) \sim R^{-5}$  for $R \rightarrow \infty$, the integrand in (\ref{Iout0})
has the following limits:
\begin{eqnarray*}
|\tilde{V}_0(u) e^{i S(u)}|  &\rightarrow& 0,      
\quad  -\frac{\pi}{4} < \theta < \frac{\pi}{4}, \frac{3\pi}{4} < \theta < \frac{5\pi}{4}, \\
|\tilde{V}_0(u) e^{i S(u)}|  &\rightarrow& \infty, 
\quad  -\frac{3\pi}{4} < \theta < -\frac{\pi}{4}, \frac{\pi}{4} < \theta < \frac{3\pi}{4}.
\end{eqnarray*}
Therefore, it is impossible to select $C_{\rm out}$ in such a way that the integrand on 
the whole contour approaches zero. Moreover, in section~\ref{sec:Test} it is 
numerically demonstrated that $I_{\rm out}$ is of the same order of magnitude
as $A(\vek{p})$ or even larger.

\subsection{Contours through steepest descent}

\begin{figure}
\includegraphics[width=80mm]{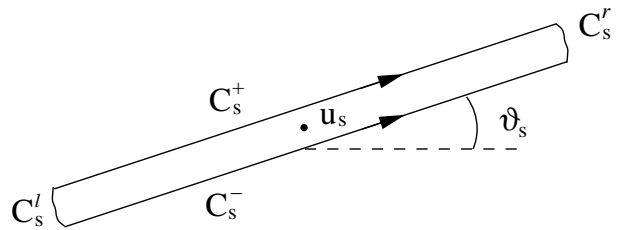}
\caption{\label{fig:C_us} The integration paths $C^{\pm}_s$ passing around  
the special point $u_s$ from both sides and given parametrically in Eq.(\ref{ux}). 
The contours $C^{\rm r}_s$ and $C^{\rm l}_s$ are used to connect the contours 
$C^{\pm}_s$ at infinity.
}
\end{figure}

In order to understand the appearance of the factor two between the RT and the
MSD results it is important to connect the two approaches. For this purpose, 
the four auxiliary integrals
\begin{equation}
I^{\pm}_s = \int_{C^{\pm}_s} g_{s}(u) \frac{e^{i S(u)}}{(u - u_{s})^{\nu}} d u
\label{Is}
\end{equation}
are introduced where the index $s=\pm$ specifies one of the two special
points. The contours $C^{\pm}_s$ are given parametrically by
\begin{equation}
 u_x = u_{s} + (x \pm i \varepsilon) \: Q_s,\quad 
                       -\infty < x < \infty,\;\;\varepsilon\rightarrow 0^{+}
 \label{ux}
\end{equation}
starting at $x \rightarrow -\infty$. Here, plus (minus) corresponds to the 
contour passing slightly above (below) the special point $u_s$ (see Fig.~\ref{fig:C_us}). 
The value of $Q_s$ is chosen in such a way that the contours $C^{\pm}_s$ are passing through
the steepest descent of $i S(u)$, i.\,e.\ as
\begin{equation}
 Q_{s} = \sqrt{\frac{2 i}{S''(u_{s})}}  
\label{Qs}
\end{equation}
where the argument $Q_{s}$ satisfies $ -\pi/4 < \arg Q_{s}  < \pi/4$ 
(see Eqs.(\ref{Q_theta})). According to (\ref{limEiS}) the integrand 
in (\ref{Is}) must then exponentially decay to 0 for $x \rightarrow \pm \infty$. 
This specific choice of $Q_{s}$ allows to directly apply MSD in the following 
subsection. Here, Cauchy integration rules~\cite{gen:hass99} are employed to
deduce three useful relations:
\begin{itemize}
\item[(i)] Deforming the contour 
in $A(\vek{p})$ to pass along $C^{+}_{-}$ in positive direction, 
along $C^{-}_{+}$ in negative direction and connecting the ends of these 
contours at infinity one obtains
\begin{equation}
  A(\vek{p}) = I^{-}_{+} - I^{+}_{-} \quad .
\label{AwI}
\end{equation}
\item[(ii)] Connecting contours $C^{\pm}_s$ at infinity with the  
contours $C^l_s$ and $C^r_s$ (see Fig.~\ref{fig:C_us}) and applying the residue theorem one obtains 
\begin{equation}
 A_s = I^{-}_s - I^{+}_s \quad.
\label{res_s}
\end{equation}
\item[(iii)] Substituting (\ref{AwI}) and (\ref{res_s}) into (\ref{Ap4}) one obtains
\begin{equation}
  I_{\rm out} =  I^{-}_{-} - I^{+}_{+} \quad .
\label{Iout}
\end{equation}
\end{itemize}
Note, the use of a sufficiently small but finite positive $\varepsilon$ yields the same 
value of $I^{\pm}_s$. This is used to compute $I^{\pm}_s$ numerically.

Keeping in mind that equations (\ref{AwI}), (\ref{res_s}), and (\ref{Iout}) are
exact and no approximations have been done so far, we apply now MSD to approximate $I^{\pm}_s$.

\begin{table*}[!]
\caption{ 
\label{tab:NumTest}
Contour integrals $I_{\pm}^{+}, I_{\pm}^{-}$ [Eq.(\ref{Is})], 
$I_{\rm out }$ [Eq.(\ref{Iout})], quantities $I_{\pm}$ [Eq.(\ref{IsMSD})], 
and  $A_{+}+A_{-}$ [Eq.(\ref{Apm})] for different principal quantum numbers $n$ 
and fixed parameters ($F = 0.02$ a.\,u., $\omega = 0.01$ a.\,u., $E_{\rm b} = 0.5$ a.\,u., 
$\hat{\vek{F}}\cdot \hat{\vek{p}} = 0.9$, $N=161$).
The exact value $A_{\rm ex}$ for the amplitude $A(\vek{p})$ [Eq.(\ref{Ap2})] is compared with  
the prediction of the simple MSD formula $A_{\rm MSD}$ [Eq.(\ref{MSD})], 
the corrected MSD formula $A_{\rm cMSD}$ [Eq.(\ref{cMSD})], and the amplitude $A_{\rm KM}$ (with appropriate
phase normalization) given by the ``two-term saddle-point approximation''~\cite{sfa:kjel06}.} 
\begin{ruledtabular}
\begin{tabular}{lrrrrr}
  & $ n = 1,\;\times 10^{-8}$ & $ n = 2,\;\times 10^{-7} $ & $ n = 3,\;\times 10^{-6} $ & 
$ n = 4,\;\times 10^{-5} $ & $ n = 5,\;\times 10^{-4} $  \\
\hline
$I_{+}^{+}$    & $-0.212 - 1.560\, i$ & $-0.320 - 1.760\, i$ & $-0.374 - 1.652\, i$ 
               & $-0.371 - 1.370\, i$ & $-0.326 - 1.034\, i$ \\
$I_{+}^{-}$    & $ 0.226 + 1.883\, i$ & $-0.302 - 1.884\, i$ & $ 0.274 + 1.347\, i$ 
               & $-0.189 - 0.758\, i$ & $ 0.105 + 0.352\, i$ \\
$I_{+}$        & $-0.219 - 1.722\, i$ & $-0.312 - 1.812\, i$ & $-0.325 - 1.495\, i$ 
               & $-0.276 - 1.045\, i$ & $-0.201 - 0.644\, i$ \\
 &&&&& \\
$I_{-}^{+}$    & $-1.180 - 1.485\, i$ & $ 1.127 + 1.539\, i$ & $-0.765 - 1.142\, i$ 
               & $ 0.406 + 0.667\, i$ & $-0.176 - 0.322\, i$ \\
$I_{-}^{-}$    & $ 0.960 + 1.248\, i$ & $ 1.027 + 1.465\, i$ & $ 0.912 + 1.428\, i$ 
               & $ 0.713 + 1.227\, i$ & $ 0.506 + 0.959\, i$ \\
$I_{-}$        & $-1.070 - 1.366\, i$ & $ 1.069 + 1.496\, i$ & $-0.835 - 1.282\, i$
               & $ 0.549 + 0.931\, i$ & $-0.317 - 0.595\, i$ \\
 &&&& \\
$A_{-}+A_{+}$  & $ 2.578 + 6.175\, i$ & $-0.083 - 0.199\, i$ & $ 2.325 + 5.569\, i$
               & $ 0.490 + 1.172\, i$ & $ 1.113 + 2.666\, i$ \\
$I_{\rm out}$  & $ 1.172 + 2.808\, i$ & $ 1.346 + 3.224\, i$ & $ 1.286 + 3.080\, i$
               & $ 1.084 + 2.597\, i$ & $ 0.832 + 1.993\, i$ \\
 &&&& \\
$A_{\rm MSD}$  & $ 1.289 + 3.088\, i$ & $-1.381 - 3.308\, i$ & $ 1.160 + 2.778\, i$
               & $-0.825 - 1.976\, i$ & $ 0.517 + 1.239\, i$ \\ 
$A_{\rm cMSD}$ & $ 1.406 + 3.367\, i$ & $-1.432 - 3.431\, i$ & $ 1.047 + 2.506\, i$                  
               & $-0.603 - 1.444\, i$ & $ 0.287 + 0.687\, i$ \\
$A_{\rm ex}$   & $ 1.406 + 3.368\, i$ & $-1.429 - 3.423\, i$ & $ 1.039 + 2.489\, i$
               & $-0.595 - 1.425\, i$ & $ 0.281 + 0.673\, i$ \\ 
$A_{\rm KM}$   & $ 1.378 + 3.300\, i$ & $-1.680 - 4.025\, i$ & $ 1.641 + 3.932\, i$ 
               & $-1.365 - 3.270\, i$ & $ 0.999 + 2.393\, i$          
\end{tabular}
\end{ruledtabular}
\end{table*}

\subsection{\label{subsec:MSD}The method of steepest descent in the presence of a singularity}

Since no difference is made between two different contour integrations around
the same saddle point (as, e.\,g., for $I^{+}_s$ and $I^{-}_s$)
in~\cite{sfa:grib97} (Appendix B), we shortly repeat the main steps. 
From (\ref{ux}) the relations 
\begin{equation}
d u = Q_s d x,\quad
\frac{1}{(u_x - u_{s})^{\nu}} = 
\frac{(\pm 1)^{\nu} }{(i Q_s)^{\nu}}\frac{1}{(\epsilon \mp i x)^{\nu}}
\end{equation}
follow. We expect the vicinity of $u_s$ to give the main contribution to the
integral and assume that $g_s(u)$ is a slowly varying function in the vicinity 
of $u_s$. Then, using the approximation
\begin{equation}
 g_{s}(u) e^{i S(u_x)} \approx  
g_{s}(u_{s}) e^{i S(u_{s})} e^{-x^2 \mp 2 i \epsilon x + \epsilon^2 }
\end{equation}
and the identity
\begin{equation}
\frac{1}{(a \mp i b)^{\nu}} = 
\frac{1}{\Gamma(\nu)} \int\limits_0^{\infty} d \eta \eta^{\nu -1} 
e^{- \eta a}e^{\pm i \eta b},\quad a>0
\end{equation}
one obtains  
\begin{eqnarray}
\nonumber
  I^{\pm}_s &\approx&  (\pm 1)^{\nu}  
\frac{g_{s}(u_{s}) e^{i S(u_{s})}}{i^{\nu} Q_s^{\nu-1}\Gamma(\nu)}   
\int\limits_0^{\infty} d \eta\, \eta^{\nu -1} e^{-\eta\epsilon + \epsilon^2} 
  \\
&\times&
\int\limits_{-\infty}^{\infty}d x\, 
 e^{-x^2   \pm i (\eta - 2 \epsilon)  x}.
\end{eqnarray}
The integration over $x$ and $\eta$ yields
\begin{equation}
 I^{\pm}_s \approx (\pm 1)^{\nu} I_s, \qquad
 I_s =   \frac{\pi g_{s}(u_{s}) e^{i S(u_{s})} }{i^{\nu}  Q_s^{\nu-1}\Gamma(\frac{\nu+1}{2})} .
\label{IsMSD}
\end{equation}
Therefore, MSD predicts $I^{\pm}_s$ to be equal for even $\nu$ and to 
differ only by the sign for odd $\nu$. Using equations (\ref{AwI}) and (\ref{Iout}) 
this result can be rewritten as
\begin{equation}
  I_{\rm out} =  (- 1)^{\nu-1} A(\vek{p})\quad\mbox{[within MSD].} 
\label{Iout2}
\end{equation}
Using (\ref{AwI}) the prediction of MSD for $A(\vek{p})$ is 
\begin{equation}
 A_{\rm MSD} = (- 1)^{\nu} I_{+} -  I_{-} \quad.
\label{MSD}
\end{equation}
Substitution of (\ref{Iout2}) into (\ref{Ap4}) shows that for odd $\nu$ MSD yields
\begin{equation}
 A(\vek{p}) \approx   (A_{+} + A_{-})/2 \qquad \mbox{[MSD, odd $\nu$].}
\label{Ap5}
\end{equation}
Its value is thus two times smaller than the one obtained with the assumption 
$I_{\rm out} = 0$. For even $\nu$ MSD predicts
\begin{equation}
 | A(\vek{p}) | \gg |(A_{+} + A_{-})|  \qquad \mbox{[MSD, even $\nu$].}
\label{Ap6}
\end{equation} 
Note, (\ref{Ap5}) and (\ref{Ap6}) are valid for every $\tilde{V}_0(u)$ 
satisfying (\ref{Vo_rep}), if $g_{\pm}(u)$ is a slowly varying function 
in the vicinity of $u_{\pm}$. The fact that for the 1s state of hydrogen-like
atoms ($\nu=3$) one finds exactly a factor 2 difference between MSD and RT and
thus an equality sign in (\ref{Ap5}) should be seen as an accidental case 
that is due to the relative simplicity of $\tilde{V}_0(u)$ for the 1s state.

In section~\ref{sec:Test} MSD is tested numerically and it is shown that the 
assumption of a slowly varying function $g_{\pm}(u)$ is valid only for small 
$n$ (or $\nu$).

\section{\label{sec:Test}Numerical test} 

\begin{figure}[!]
\includegraphics[width=80mm]{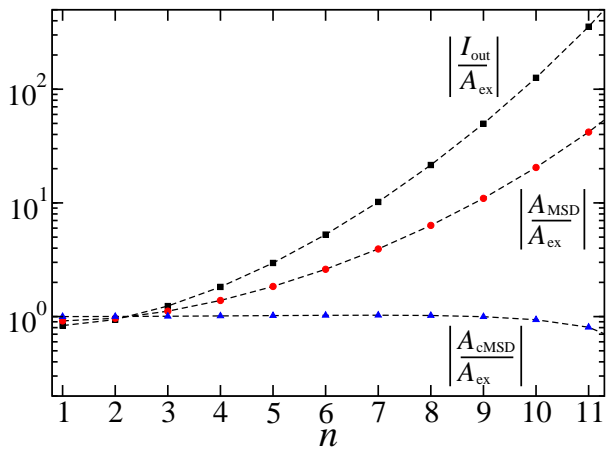}
\caption{\label{fig:Ratio}  
(Color online) Demonstration of the failure of the MSD for large principal quantum numbers
$n$. The ratio $|I_{\rm out}/A_{\rm ex}|$ (boxes) grows
exponentially with $n$ which causes an increase of the ratio  
$|A_{\rm MSD}/A_{\rm ex}|$ (circles) between the approximate and exact
amplitudes $A_{\rm MSD}$ and $A_{\rm ex}$, respectively. Therefore, the
simple (standard) MSD formula [Eq.(\ref{MSD})] fails and must be improved to
be applicable for large $n$. The in this work proposed corrected MSD formula 
[Eq.(\ref{cMSD})] shows very good accuracy in a large region of $n$, as can be
seen from the corresponding ratio $|A_{\rm cMSD}/A_{\rm ex}|$ (triangles).
The parameters used in the computation are the same as in Table~\ref{tab:NumTest}.
}
\end{figure}

To support our conclusions of the previous section the example results of a 
numerical study are reported in Table~\ref{tab:NumTest}. For a more
transparent analysis a number of parameters were fixed. This includes the
amplitude of the electric field $\vek{F} = 0.02$ a.u., the frequency $\omega = 0.01$ a.u.,
the binding energy $E_{\rm b} = Z^2/(2 n^2) = 0.5$ a.u.\ (thus charge $Z = n$), 
the angle $\hat{\vek{F}}\cdot \hat{\vek{p}} = 0.9$,
and the number of photons $N=161$. With such a choice of fixed parameters a
variation of the principal quantum number $n$ leaves the function $S(u)$
unchanged (see Appendix A for details). The same is true for the positions of the 
special points $u_{\pm}$ and the values of $Q_{\pm}$. Therefore, only  
function $\tilde{V}_0(u)$ varies with $n$. 

We use (\ref{Ap2}) to calculate the exact value $A_{\rm ex}$ of $A(\vek{p})$
and (\ref{Apm}) to evaluate $A_{\pm}$. The integrals $I^{\pm}_s$ are calculated numerically, 
Eq.(\ref{AwI}) and (\ref{res_s}) are used for a check of the numerics and  
$I_{\rm out}$ is obtained from (\ref{Iout}).

As follows from the discussion above, the condition $I^{-}_s \approx
(-1)^{\nu} I^{+}_s$ obtained in (\ref{IsMSD}) can be used 
as a criterion for the validity of the simple MSD formula (\ref{MSD}).
Table~\ref{tab:NumTest} shows that this condition is fulfilled for $n=1,2$ 
and the relations (\ref{Iout2},\ref{Ap5},\ref{Ap6}) are valid. With increasing  
$n$ the condition is, however, not well fulfilled and the accuracy of the 
MSD prediction decreases. Numerical tests show that for large $n$ 
the values $|I^{+}_{+}|,|I^{-}_{-}|$ are by orders of magnitude larger
than $|I^{-}_{+}|,|I^{+}_{-}|$. This leads to the following relations 
(see Fig.\,\ref{fig:Ratio})
\begin{equation*}
|A_{\rm ex}| \ll |A_{\rm MSD}|, \quad 
|A_{\rm ex}|  \ll |I_{\rm out}|,\quad I_{\rm out} \approx  A_{+} + A_{-}. 
\end{equation*}
Therefore, function $\tilde{V}_0(u)$ cannot be given as simple as 
in (\ref{Vo_rep}). Instead, $\tilde{V}_0(u)$ can be represented in the vicinity 
of $u_s$ by a sum over terms having different orders of poles (see Eq.(\ref{V_0u2})).
Moreover, it is possible to consider also higher derivatives of $S(u)$, as
is done in Eq.(\ref{eSuexp}). 
The resulting representation of $\tilde{V}_0(u)\exp[i S(u)]$ given in Eq.(\ref{VoESrep})
and the subsequent use of MSD to it leads to a much higher accuracy. 
As can be seen from Table~\ref{tab:NumTest} and  Fig.\,\ref{fig:Ratio} the
corrected MSD formula (\ref{cMSD}) yields a significant improvement
and can be used for the numerical computation of $A(\vek{p})$ in a large
range of $n$. This paves the way for a detailed study of the validity 
of the Keldysh approximation for, e.\,g., Rydberg atoms that is not 
blurred by a failure of the usually adopted MSD approximation.

It is instructive to compare $A_{\rm cMSD}$ with the recently 
published ``two-term saddle-point approximation''~\cite{sfa:kjel06}. 
This approximation is a modification of the 
``one-term  saddle-point approximation''~\cite{sfa:grib97}. 
Both approximations are based on the principle that only the leading term of 
the Laurent expansion of the Fourier transform $\tilde{\Phi}_{0}(\vek{q})$ 
at the saddle points is considered. Since in the ``one-term'' approximation 
the contributions from higher-order derivatives of $S(u)$ are also ignored, 
it is equivalent to the simple MSD formula $A_{\rm MSD}$ and yields thus the
same numerical results. (Note, the resulting expressions are, however, different,
because different complex variables $\phi= \omega t$ and $u$ are used in the 
derivations.) As can be seen from Table~\ref{tab:NumTest}, 
the amplitude $A_{\rm KM}$ yielded by the ``two-term'' approximation (using an 
appropriate phase normalization) gives relatively good agreement for $n=1$, although 
it is clearly less accurate than  $A_{\rm cMSD}$. For the excited states the 
``two-term'' formula fails rapidly and yields results which are even less
accurate than the ones obtained with the ``one-term'' formula. To understand
this fact, we remind that the second term in~\cite{sfa:kjel06} comes from the 
next-to-leading pole, and only the contribution arising from the third derivative of
$S(u)$ is considered. A detailed analysis shows that a singularity of the
same order can also arise from  the next-to-leading term of the Laurent
expansion of $\tilde{\Phi}_{0}(\vek{q})$. In addition, for larger values of
$n$ it can also be important to consider further terms of the Laurent
expansion. Since the corrected MSD formula (\ref{cMSD}) takes all of this into 
account, the resulting $A_{\rm cMSD}$ is significantly more accurate than 
$A_{\rm KM}$.

\section{Conclusion}

In this work it has been demonstrated that the residue theorem was not 
correctly employed in~\cite{sfa:mish02,sfa:mish02b}, since the derivation 
was based on an unjustified assumption that one integral vanishes.  
This neglected term is, however, of the same 
order of magnitude as the remaining ones or even much larger. For the 1s state 
of hydrogen-like atoms it is almost identical, and thus its omission results in an 
overestimation of the transition amplitude by a factor two 
for this case. This deviation was in~\cite{sfa:mish02} incorrectly 
assumed to be a failure of the widely used saddle-point approximation. 
Considering a 2s state, it is furthermore concluded that an application 
of the method proposed in~\cite{sfa:mish02} to a 2s state would yield an 
even larger (erroneous) deviation. 

Such a large deviation for the 2s state was in fact reported 
in~\cite{sfa:chao05} where also the residue method had been applied. 
Analogously to~\cite{sfa:mish02} the deviation was attributed to a failure 
of the saddle-point method, but is in fact due to the same unjustified omission 
of a non-vanishing integral. The direct applicability of the present findings 
to~\cite{sfa:chao05} can be verified, since the derivation 
in~\cite{sfa:chao05} differs from~\cite{sfa:mish02} essentially only by the 
choice of $\phi = \omega t$ as complex variable, while in~\cite{sfa:mish02} 
and the present work $u = \sin\omega t$  was used.    

The applicability of the method of steepest descent (saddle-point
approximation) for arbitrary $n$s states has also been investigated in the present 
work. It is found that the simple standard formula fails for large $n$. 
To overcome this problem a corrected formula is proposed. 

\section*{Acknowledgments}
AS and YV acknowledge financial support by the {\it Deutsche 
Forschungsgemeinschaft}. AS is grateful 
to the {\it Stifterverband f\"ur die Deutsche Wissenschaft} (Programme  
{\it Forschungsdozenturen}) and the {\it Fonds der Chemischen Industrie} 
for financial support. 

\appendix

\section{\label{app:Sp} Calculation of $u_{\pm}$, $S(u),S'(u),S''(u_{\pm}),S'''(u_{\pm})$, 
 and $Q_{\pm}$.}

The integration of (\ref{Su}) for $|\vek{p}| = p_N$ yields
\begin{eqnarray}
\nonumber
 \exp[i S(u)] &=& \exp\left\{ i \frac{\vek{p}\cdot\vek{F}}{\omega^2}[1-f(u)] 
- i \frac{U_p}{\omega} u f(u)\right\} \\
&\times& [ f(u) + i u]^N  \quad  .
\label{expiSu}
\end{eqnarray}
For $u = R e^{i\theta}$ with $R \rightarrow \infty$, one has
$f(u) \rightarrow - i u + i u^{-1}/2$ and
\begin{equation}
 \exp[i S(u)] \rightarrow   \frac{\exp\left\{ -  \frac{U_p}{\omega} u^2 
- \frac{\vek{p}\cdot\vek{F}}{\omega^2} (u-i) \right\} }{ (-2 i u)^{N}}
\label{expiSu_as}.
\end{equation}

The saddle points $u_{\pm}$ of $S(u)$ can be determined by 
the following condition:
\begin{equation}
\frac{E_{\rm b}}{\omega} +  \frac{1}{2\omega}\left(\vek{p} + 
\frac{\vek{F}}{\omega} u \right)^2 = 0.
\end{equation}
Introducing the Keldysh parameter $\gamma = \kappa \omega/F$ with 
$\kappa = \sqrt{2 E_{\rm b}}$, the scaled momentum $\chi = p_N/\kappa$, and
$\zeta = \hat{\vek{F}}\cdot \hat{\vek{p}}$
\begin{equation}
 u_{\pm} = -\sigma \pm \rho i,\quad \sigma = \gamma \chi \zeta,\; \rho = 
\gamma\sqrt{1 + \chi^2(1 - \zeta^2)}
\label{upm}
\end{equation}
is obtained. Using (\ref{upm}) the first derivative $S'(u)$ can be expressed as
\begin{equation}
 S'(u) = \frac{2 U_p}{\omega} \frac{( u - u_{+})(u - u_{-})}{f(u)}
\end{equation}
and the values of the second $S''(u)$ and third $S'''(u)$ derivatives at 
$u = u_{\pm}$ are given by
\begin{equation}
  S''(u_{\pm}) = \pm  \frac{4 U_p \rho i}{\omega f(u_{\pm})}, \quad
  S'''(u_{\pm}) =  \frac{6 U_p - 2 N \omega}{\omega f^3(u_{\pm})}. 
\end{equation}

The absolute value $Q$ and the argument $\vartheta_s$ of $Q_s$ defined 
by (\ref{Qs}) can be written as
\begin{equation}
Q = \frac{\sqrt{2\omega}\gamma}{\sqrt{\rho} \kappa}  
[(1+\sigma^2 + \rho^2)^2 - 4\sigma^2]^{1/8},
\end{equation}
\begin{equation}
\tan 4 \vartheta_{\pm} = \pm \frac{ 2 \sigma \rho}{ 1 + \rho^2 - \sigma^2},\quad 
 -\frac{\pi}{4} < \vartheta_{\pm} < \frac{\pi}{4}.
\label{Q_theta}
\end{equation}

Note, for the small momentum limit $\vek{p} \ll \kappa$ the following
relations are valid:
\begin{equation}
\rho \approx \gamma,\quad 1 - u^2_{\pm} 
\approx 1 + \gamma^2 \pm 2 \gamma^2 \chi \zeta i,
\end{equation}
\begin{equation}
Q \approx \frac{\sqrt{2\omega\gamma}}{\kappa}(1+\gamma^2)^{1/4},\quad 
\vartheta_{\pm} \approx 
\pm \frac{ \chi \zeta\gamma^2}{2(1+\gamma^2)}.
\end{equation}

\section{\label{app:Vo} Function $\tilde{V}_0(u)$ for the $n$s states 
of a hydrogen-like atom.}

Consider the spherically symmetric state (with principal quantum number $n$)
of a hydrogen-like atom with potential $U(\vek{r}) = Z/r$, where 
$Z$ is the charge of the nucleus. Its Fourier transform is given by
\begin{equation}
\tilde{\Phi}_{0}(\vek{q}) = \frac{8\sqrt{\pi}}{\kappa^{3/2}} 
\sum_{k=0}^{n-1} (-1)^k 2^{2k} C^{n+k}_{2k+1} 
\left(\frac{ \kappa^2}{\vek{q}^2 + \kappa^2}\right)^{k+2}
\end{equation}
where $C^n_k$ are binomial coefficients.
Using the identity $\nabla_{\vek{q}} f(q^2) = 2 \vek{q}  
\partial f(q^2)/( \partial q^2)$ one can rewrite Eq.\,(\ref{Vo}) as
\begin{equation}
V_{0}(\vek{q}) = \frac{4\pi(\vek{F}\cdot\vek{q})}{\kappa F} 
\sum_{k=3}^{n+2} D_n^{(k)} 
\left(\frac{ \kappa^2}{\vek{q}^2 + \kappa^2}\right)^{k}
\end{equation}
where
\begin{equation}
D_n^{(k)} =   (-1)^{k}\, (k-1)\, 2^{2k-4}\, 
C^{n+k-3}_{2k-5}\frac{i F}{\sqrt{\pi}\kappa^{5/2} }.
\end{equation}
Introducing 
\begin{equation}
P_{\pm}(u) = \frac{\gamma}{u - u_{\pm}}
\end{equation}
and using
\begin{equation*}
\vek{F}\cdot\vek{q} = \frac{F \kappa}{2}[ P^{-1}_{+}(u) + P^{-1}_{-}(u)],\quad
\frac{ \kappa^2}{\vek{q}^2 + \kappa^2} = P_{+}(u)  P_{-}(u),
\end{equation*}
one can rewrite  Eq.\,(\ref{Vou})  as
\begin{equation}
  \tilde{V}_0(u) = 
\sum_{k=3}^{n+2} D_n^{(k)} \{ P^{k-1}_{+} P^{k}_{-} + P^{k}_{+} P^{k-1}_{-} \}.
\label{V_0u}
\end{equation}
Introducing $R_{\pm} = P_{\pm}(u_{\mp}) = \mp \gamma/(2 \rho i)$ and using the 
Taylor expansion of $P^k_{\pm}(u)$ at $u \approx u_{\mp}$,
\begin{equation}
   P^k_{\pm}(u) = \sum_{m=0}^{\infty} (-1)^m C^{k+m-1}_m R^{k+m}_{\pm} P^{-m}_{\mp}(u),
\end{equation}
one can rewrite $\tilde{V}_0(u)$ as a Laurent series at $u \approx u_{\pm}$
\begin{equation}
   \tilde{V}_0(u) = \sum_{\nu=-\infty}^{n+2} \frac{ M^{(n,\nu)}_{\pm}}{(u - u_{\pm})^\nu} 
\label{V_0u2}
\end{equation}
where
\begin{equation}
   M^{(n,\nu)}_{\pm} = \gamma^{\nu} \hspace{-3mm}\sum_{r=\max(\nu,3)}^{n+2} 
                       D_n^{(r)} Q^{(r)}_{r-\nu} R_{\mp}^{2r-\nu-1},
\end{equation}
and
\begin{equation}
   Q^{(k)}_{m}  = (-1)^m \{ C^{k+m-2}_m - C^{k+m-2}_{m-1}\},\quad Q^{(k)}_{0} = 1.
\end{equation}
Then for $g_{\pm}(u)$ defined in (\ref{Vo_rep}),
\begin{equation}
  g_{\pm}(u) = \sum_{m=0}^{\infty} M_{\pm}^{(n,n+2-m)} (u - u_{\pm})^m,
\end{equation}
 one has
\begin{equation}
  g_{\pm}(u_{\pm}) = M_{\pm}^{(n,n+2)} = (\pm 2)^{n-1} \frac{i^{n} (n+1)
                     \gamma^{2n+3} F}{\sqrt{\pi}\rho^{n+1}\kappa^{5/2}}.
\label{gsus}
\end{equation}

\section{\label{app:cMSD} Corrected MSD formula for the $n$s states of 
a hydrogen-like atom.}
Representing $\exp[i S(u)]$ as
\begin{eqnarray}
\nonumber
  \exp[i S(u)] &=& \exp[i S(u) - (i/2) S''(u_s) (u - u_s)^2] \\
            &\times& \exp[ - Q^2_s (u - u_s)^2]
\label{eiSurep}
\end{eqnarray}
and performing a Taylor expansion of the first term on the right hand side 
of (\ref{eiSurep}) at $u=u_{s}$ yields
\begin{eqnarray}
\nonumber
e^{i S(u)} &=& e^{i S(u_{s})} e^{-Q^2_s (u - u_s)^2 } \\
&\times& \left\{ 1 + \frac{i S'''(u_s)}{6}(u-u_s)^3+ \dots\right\}. 
\label{eSuexp}
\end{eqnarray}
Keeping the first two terms of the Taylor expansion the integrand 
in (\ref{Ap3},\ref{Iout0}) can be rewritten as
\begin{eqnarray}
\nonumber
  \tilde{V}_0(u) e^{i S(u)} &\approx& e^{i S(u_{s})} e^{-Q^2_s (u - u_s)^2 } 
\left\{ \sum_{\nu=-\infty}^{n+2} \frac{ M^{(n,\nu)}_{s}}{(u - u_{s})^\nu}  \right. \\
  &+& \frac{i S'''(u_s)}{6} \left. \sum_{\nu=-\infty}^{n-1} 
\frac{ M^{(n,\nu+3)}_{s}}{(u - u_{s})^\nu}
\right\}.
\label{VoESrep}
\end{eqnarray}
Omitting terms with negative $\nu$ in (\ref{VoESrep}) and applying 
the procedure described in Sec.\,\ref{subsec:MSD} 
 \begin{eqnarray}
\nonumber
  I^{\pm}_{s;cMSD} &=& \pi e^{i S(u_{s})}  
\left\{ \sum_{\nu=0}^{n+2} \frac{ (\pm 1)^{\nu} M^{(n,\nu)}_{s}}{i^\nu Q^{\nu-1}_s 
\Gamma\left(\frac{\nu+1}{2} \right)}  \right. \\
  &+& \frac{i S'''(u_s)}{6} \left. \sum_{\nu=0}^{n-1} \frac{ (\pm 1)^{\nu} M^{(n,\nu+3)}_{s}}
{i^\nu Q^{\nu-1}_s \Gamma\left(\frac{\nu+1}{2} \right)}
\right\}
\label{IscMSD}
\end{eqnarray}
is obtained as approximation for $I^{\pm}_{s}$. 
Note, that neglecting all terms with $\nu<n+2$ in (\ref{IscMSD}) and 
using (\ref{gsus}) one obtains the simple (standard) MSD formula (\ref{IsMSD}) 
for $I^{\pm}_{s}$. A corrected approximation for $A(\vek{p})$ is then obtained 
using (\ref{AwI}) as 
\begin{equation}
   A_{cMSD}  = I^{-}_{+;cMSD} - I^{+}_{-;cMSD}.
\label{cMSD}
\end{equation}

\bibliographystyle{apsrev} 

\end{document}